# Santa and the moon

Peter Barthel
Kapteyn Astronomical Institute,
University of Groningen,
The Netherlands
*pdb@astro.rug.nl*

Abstract
Happy end-of-the-year evening and night events provide good opportunities to explain the phases of the moon. The need for such moon phase education is once again demonstrated, through an investigation of illustrations on Santa Claus and Christmas gift wrap and in children's books, in two countries which have been important in shaping the image of Santa Claus and his predecessor Sinterklaas: The Netherlands and the USA. The moon on Halloween illustrations is also considered. The lack of knowledge concerning the physical origin of the moon phases, or lack of interest in understanding, is found to be widespread in The Netherlands but is also clearly present in the USA, and is quite possibly global. Definitely incomplete, but surely representative lists compiling both scientifically correct and scientifically incorrect gift wrap and children's books are also presented.

I. Introduction
Indicating an evening or night scene, images of the moon are often used, not only as illustrations in books (for adults and children), but also on product packings, brochures, greetings cards, gift wrap, in advertizing, commercials, as pictogram, etcetera. A full moon is often used, partly hidden behind trees or clouds, but also a partially lit moon - half or crescent - is frequently seen. The latter leaves no doubt as to identification with the moon; a full moon could possibly be mistaken for the sun. As is well known there are two possibilities to depict a partially lit moon: around its first, and around its last quarter. A crescent moon, on its way towards first quarter is called a waxing moon, and such a waxing moon can be observed in afternoon twilight and in the evening. Its right hand side is lit, for observers in the northern hemisphere (left for observers in the south). The first quarter moon sets at midnight, a waxing moon even earlier. A waning moon crescent (from third quarter to new moon: left hand side lit) rises around 3am, hence can only be observed in late night and in the morning twilight. Full moon is directly opposite the sun in the sky, hence rises at sunset.

Moon illustrations are occasionally incorrect. Post cards exist showing artist impressions of tropical evenings with sun and full moon close to each other. Full moons sometimes rise around midnight in movie scenes. Illustrations which show moon crescents are also occasionally wrong: third quarter moons or waning crescents are being depicted when the actual scene is in the evening. A 2010 Unicef Christmas card as well as the opening scene in the 2010 Jacquie Lawson animated Advent e-Calendar - widely sold items - provided the culmination of several years of frowning (and smiling) about this misconception or ignorance, and triggered the research presented below. The Unicef card, of British design showed children decorating an outdoor Christmas tree. Judging from the moon phase, the scene takes place at 4am or 5am in the morning, which is not impossible but unlikely. The village scene which opened the 2010 Jacquie Lawson (http://www.jacquielawson.com) digital Advent Calendar depicts a Christmas caroling event, with performers and listeners including church choir members, on a village square. Whereas the thin waning moon indicates an early morning event, the artist undoubtedly wants to show us an early evening scene, judging from the people on the square and the lights in the houses, stores, and church. One cannot exclude the possibility that both artists had the intention to create Australia scenes with reversed moon phases, but the presence of snow in both scenes is strongly suggestive of the Northern Hemisphere December month … The same misunderstanding (or ignorance) is frequently seen on gift wrap in The Netherlands, displaying Sinterklaas, the predecessor of Santa Claus, distributing presents in the evening with a third-quarter or waning moon in the sky.

To quantify the level of ignorance concerning the phase of the evening moon related to the Sinterklaas, Santa Claus and Christmas season, a (jolly) research project was conducted, examining illustrations in children's books, on gift wrap, and on Christmas cards, in the USA and The Netherlands. These are the two countries which shaped the image of Santa Claus and his name giver Sinterklaas (Saint Nicholas, Sint Nicolaas), the benevolent figures that have been of key importance in commercializing the December month. Goal of publishing this research is obviously to focus educators' attention on the great opportunity to explain the (origin of the) moon phases, as offered by the happy end-of-the-year events. The need for such education has been for instance well demonstrated by the video "A Private Universe" (http://www.learner.org/resources/series28.html)

II. Theoretical background
The Sinterklaas and Santa Claus stories are well known and date back to the fourth century bishop Nicholas of Myra (now Demre, in south-Turkey). Legends of his generosity and kindness spread over Europe and he became the patron saint of many groups, cities, and even countries. December 5, the eve of his death, was commemorated with an annual feast in many European countries. Following the protestant reformation in the sixteenth century, some countries merged the St.Nicholas celebration with Christmas but others stuck to December 5. Its character of benevolence and the exchange of gifts obviously draws back to legends around Nicholas. Local folklore was added which resulted in different flavours for the celebrations in various countries. Dutch settlers in the New World celebrated St.Nicholas eve in the 17th century, and those celebrations evolved through contacts with immigrants of other nationalities. Concerning the transformation to Santa Claus, the importance of author Washington Irving in 1809 and of the 1823 poem "A visit from Saint Nicholas", commonly known as "The night before Christmas"[1] is well documented, as are the drawings of Thomas Nast in the 1860's and the Coca-Cola advertisements in the 1930's (e.g., Jones 1978, Crichton 1987 and http://www.stnicholascenter.org).

The Dutch Sinterklaas together with his helpers is believed to distribute his presents on the evening of December 5. He may also pay visits during the evenings and nights before December 5, leaving some candy in the shoes before the fireplace, but the main event is the evening just mentioned. Once delivered, the presents are unwrapped during that same evening. Scenes depicting the December 5 Sinterklaas events should depict waxing, first quarter or full moon (or no moon) to be scientifically correct.

---

[1] The widely known poem, with credit to Clement Clarke Moore, can be found for instance on http://poets.org/viewmedia.php/prmMID/19286. Site http://rpo.library.utoronto.ca/poem/1312.html attributes it to Henry Livingston, Jr., following research published in 2000 by Donald Foster. A critical assessment of the authorship case can be read on http://www.common-place.org/vol-01/no-02/moore/

Santa Claus on the other hand works all night of December 24-25 to get his job done: the gifts are being unwrapped on Christmas morning. If a moon is shown in a Santa scene taking place during the beginning of the night, its phase should be waxing, quarter or full, otherwise it can have all phases. The "A visit from Saint Nicholas" poem has the family father witnessing Santa's arrival before going to bed, thus in the evening. The moon "on the breast of the new fallen snow" must have been a waxing, first quarter, or full moon to be scientifically correct.

In summary, illustrations should preferentially show first quarter or full moons on Sinterklaas or Santa Claus scenes, unless it is clear from the text that the latter man's job is nearly done (nearly empty sleigh…) Any winter evening scene in general, and certainly those having children around, should have a first quarter or full moon, if the artist wants to show one.

III. Measurements and results

III-1. December events

All data were obtained in the months November 2010 - January 2011, and can be found compiled in Tables I through IV, which are available on-line[2]. In The Netherlands, two dozen book stores and department stores were visited, during November - December 2010. Book illustrations depicting Sinterklaas and the moon were examined, with regard to the moon phase. Twenty-five images in/on twenty-five books (Table I) provide the pie chart statistic as shown in Figure 1. It is seen that 40% of the pictures display the last quarter moon, which is incorrect.

*** Figure 1 – Netherlands book statistics ***

The Netherlands gift wrap analysis (Table II: 20 different designs, from various firms, department stores and book stores) indicates an even higher level of misunderstanding. The chart in Figure 2 indicates a 65% occurrence of the incorrect last quarter or waning moon. There is no reason to believe that this statistic is not representative, because examination of a collectioner's sample of older Sinterklaas gift wrap (http://www.sinterklaaspapier.nl) yields the comparable figure of 67% (six out of nine moon-designs).

*** Figure 2 – Netherlands gift wrap statistics ***

It must be concluded that the Dutch are often wrong, both in the gift wrap and the book illustrations, more so in the former. Figure 3 nevertheless presents an example of nice gift wrap with a correct waxing evening moon.

*** Figure 3 – Scientifically correct Sinterklaas gift wrap ***

The USA research was carried out during November 2010 - January 2011, in New England and in Los Angeles: also here roughly two dozen book stores, stationery stores, pharmacies and department stores were visited. In addition, samples of commercially available gift wrap and Christmas cards were inspected on the internet. Thirty-three moon scene images in thirty different books (Table III) provide the pie chart statistic as shown in Figure 4. It is clear that most of the time (70%) full moon is drawn, but it should be noted that seventeen of the thirty inspected books were renditions by various illustrators of the classical "A visit from Saint Nicholas" poem. Inspection of 19th and early 20th century, illustrated editions of this poem
(http://www.iment.com/maida/familytree/henry/illos/editions/index.htm)
indicates that full moon is indeed shown very frequently, throughout the lifetime of this poem. Whereas its text does not explicitly mention full moon, that moon phase may have become implicitly "attached" to its lines, throughout the years: full moon moreover provides a nice background for sleigh and reindeer ... Nevertheless, in several American books an incorrect waning or third-quarter moon is seen accompanying stories involving children in evening scenes, or illustrating Santa with a full sleigh, i.e., at the beginning of a night's work. The booklet "The Night before Christmas in California" (Smith & Egan 1992) displays two different moons (full and waning) during one and the same evening ...

---

[2] http://www.astro.rug.nl/pdb/santa

*** Figure 4 – USA book statistics ***

With only a few exceptions, gift wrap and Christmas cards sold in the USA (book stores, general and department stores, stationery stores and pharmacies: single cards and boxed sets) were found to display full moons, with or without Santa. The Unicef card referred to in the Introduction represents one of the incorrect cases. Also a 2010 Holiday Delices set (cookies), sold by a famous (Fifth Avenue) department store, was found displaying a last quarter moon at the beginning of a night of hard work by Santa and his crew. It must be concluded that the Americans are occasionally wrong, but not as frequently as the Dutch.

III-2. Other events

Other relevant happy outdoor evening events include Trick-or-treat (Halloween, the eve of Old Hallows) and Sint Maarten (Saint Martin), widely celebrated each year respectively in the USA, the UK, and Canada (October 31) and The Netherlands, Belgium, France, and Germany (November 11). Often carrying lanterns and costumed, children go from home to home, begging for candy or money. Short songs about Sint Maarten are performed, whereas Trick-or-treat is obviously accompanied by an innocent threat to the home owner's property. Given that these are early evening activities, any moon should be waxing, first quarter, or full, also in book illustrations. With reference to Table IV, inspection of eleven American Trick-or-treat books indicated five cases (45%) of an incorrect last quarter moon, one first quarter, and five full moons. Three Dutch books dealing with Sint Maarten all showed the incorrect last quarter moon. Two of these, however, were translations from German and Swiss editions: it is conceivable that the misunderstanding or lack of knowledge is global.

The well known classic children's book "Goodnight Moon" (Brown & Hurd 1947) displays a correct full (evening) moon, while the various moon phases are also correctly dealt with in the classic "Moon Man" (Ungerer 1967). On a last note, however, several titles in the Good Night Our World boardbook series (www.goodnightourworld.com), where children say goodnight to their city or region, display incorrect last quarter moons.

IV. Discussion

We have established that illustrators and designers draw moons ad libitum, according to their taste, but often physically incorrect. The most common mistake is the early morning waning moon shown in an evening scene. Our research focussed on Sinterklaas, Santa Claus, and Christmas scenes, with a short side trip to Sint Maarten and Halloween. The apparent lack of knowledge concerning the physics of the moon phases is most likely widespread and not just limited to the countries examined here. Further investigations are however outside the scope of the present research. We note in passing that also the psyche could play a role: people may for instance be more inclined to draw moon crescents which are open to the right, that is Northern hemisphere waning moons.

Naturally, the question arises: so what, who cares? The errors are innocent, somewhat comparable to incorrectly drawn rainbows, with the red colour at the inside of the arc. Now, watching beautiful natural phenomena like rainbows and moon crescents is one thing, but understanding them makes them all the lot more interesting. Moreover, understanding leads to knowledge which lasts. A tiny bit of insight that full moon, being opposite to the sun, rises when the sun sets and sets when the sun rises, that first quarter moon, being at ~$90^o$, sets around midnight, and that last quarter moon, at ~$270^o$, rises around that time, is required to avoid making the reported mistakes. The Halloween, Sint Maarten, Sinterklaas, and Santa Claus settings provide wonderful opportunities for moon phase education[3], through simple naked eye observations. If this paper stimulates that education and leads to improved understanding, we would all be pleased. "You better watch out" is often heard in the December month: it should also be taken literally ....

Acknowledgments
The author acknowledges useful comments on the manuscript by Drs. Andrew Fraknoi, Bob Sanders, Scott Trager, Piet van der Kruit, Gijs Verdoes Kleijn, and particularly three unknown referees. Travel grants from the Leiden Kerkhoven Bosscha Fund and the Kapteyn Astronomical Institute are gratefully acknowledged.

References
Brown, M.W., & Hurd, C. 1947, and later editions, *Goodnight Moon*, Harper Collins.
Crichton, R. 1987, *Who is Santa Claus? - the true story behind a living legend*, Edinburgh: Canongate.
Jones, C.W. 1978, *Saint Nicholas of Myra, Bari, and Manhattan: Biography of a Legend*, Chicago: University of Chicago Press.
Smith, C., & Egan, S. 1992, *On the night before Christmas, all down 101…*, Salt Lake City: Gibbs Smith Publ.
Ungerer, T. 1967, and later editions, *Moon Man*, Harper Collins.

---

[3] Useful general moon phase education webpages are maintained by NASA and by NOAO:
http://www.jpl.nasa.gov/education/index.cfm?page=123 and http://www.noao.edu/education/phases
Excellent moon phase activities can be found on http://www.dennisschatz.org/activities.html

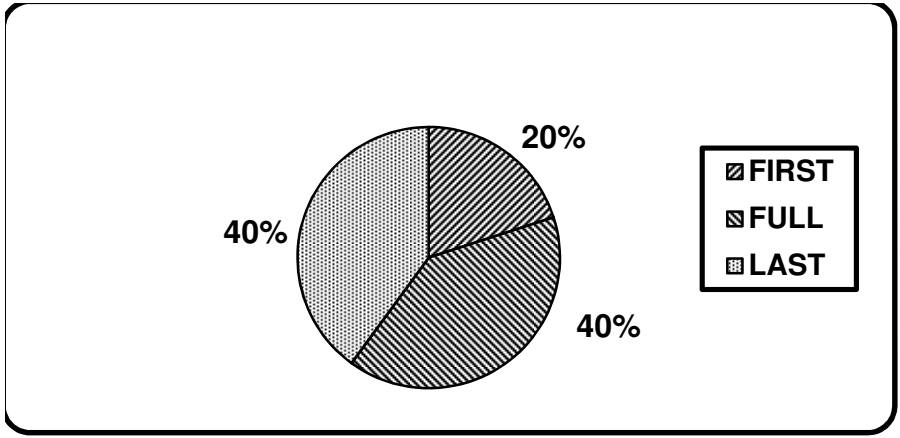

Figure 1: Netherlands book statistics

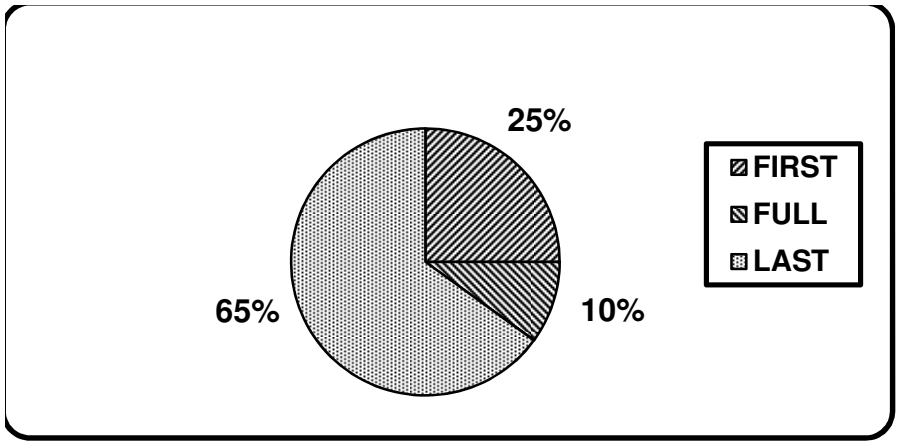

Figure 2: Netherlands gift wrap statistics

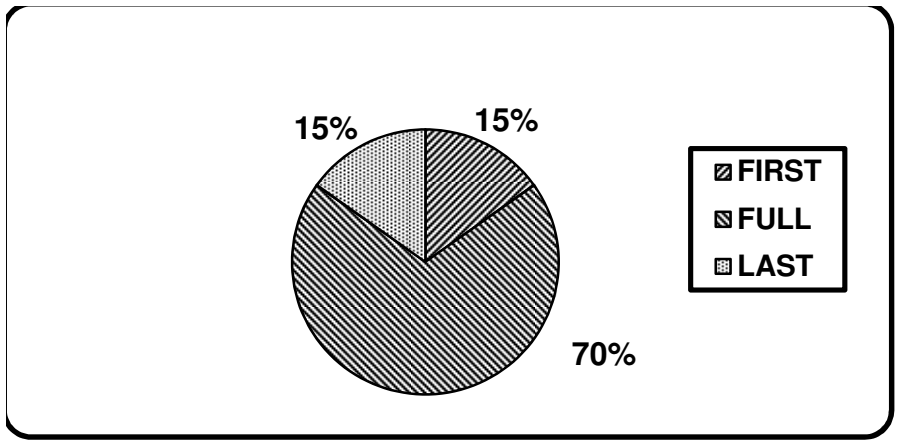

Figure 4: USA book statistics

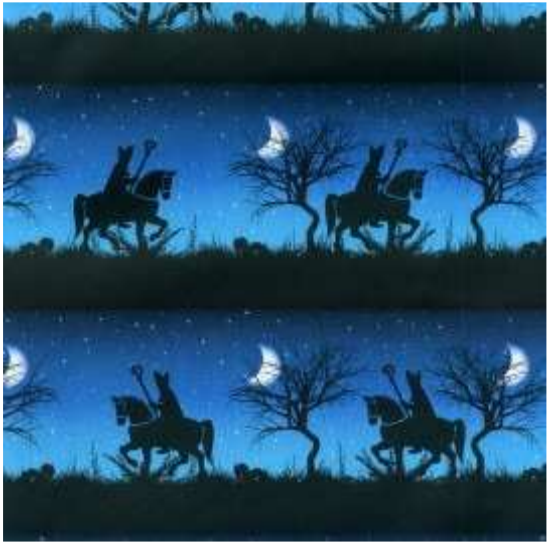

Figure 3: Scientifically correct Sinterklaas gift wrap - Keijzer Co., Wormerveer, The Netherlands: type 90011 (design reproduced with permission)

| Nr. | TABLE A - SINTERKLAAS BOOKS (NL) | Authors | Publisher | Year | Phase |
|---|---|---|---|---|---|
| 1 | Sinterklaasliedjes (+ CD) | | Mediadam | 2009 | full |
| 2 | Zou de goede Sint wel komen | N. van den Hurk | Zwijsen | 2006 | full |
| 3 | Zeppelin Sinterklaasjournaal | | omroep.nl | | first |
| 4 | Sint Siem | D. Schothorst | Zwijsen | 2005 | last |
| 5 | Sinterklaasliedjes | | BSN b.v. | | first |
| 6 | Klaas | W.G. vdHulst | Callenbach | 2008 | full |
| 7 | Groeten van de Sint | Dros & Geelen | Querido | 2008 | first |
| 8 | Lieve Sinterklaas | Amant | Clavis Peuter Serie | 2009 | last |
| 9 | Heerlijkste Sinterklaas Voorleesboek (cover) | | Querido | 2009 | last |
| 10 | Grote Sinterklaas versjes en verhalen boek (cover) | H. van Straaten | Kluitman | | first |
| 11 | Vandaag ben ik een zwarte Piet (cover) | | vHolkema&Warendorf | | first |
| 12 | Het Sinterklaasboek | Busser & Schröder | The House of Books | 2009 | last |
| 13 | Hoe Kees Piet werd | deVoogd | Moon Books | 2010 | full |
| 14 | De Leukste Sinterklaasverhalen (Deltas) | vOudheusden | Zuid-Ned.Uitgeverij | 2007 | full |
| 15 | Sinterklaas & Pierewiet | Sluyzer & Diederen | Kimio | 2009 | first |
| 16 | De mooiste Sinterklaasliedjes (Deltas) | | Zuid-Ned.Uitgeverij | 2007 | last |
| 17 | Winky en het paard van Sinterklaas (2 books - cover) | Tamara Bos | Leopold | 2010 | last |
| 18 | Slot Marsepeinstein | | Studio100 | 2010 | full |
| 19 | O kom er eens kijken | Fredriks & Rijnsburger | Ploegsma | 2010 | full |
| 20 | Kabouter Sinterklaasboek | Poortvliet | Kok, Kampen | 2001 | full |
| 21 | De kleren van Sinterklaas | Biegel & teLoo | Lemniscaat | 2009 | last |
| 22 | Roet in het eten | vanDort | Callenbach | 2009 | full |
| 23 | De echte Sinterklaas vertelt | vanSpeyk & vanderSteen | Van Goor | 2005 | last |
| 24 | Dikkie Dik viert Sinterklaas | Boeke | Gottmer | 2009 | last |
| 25 | Sinterklaas is jarig | Nans van Leeuwen | Hema | ~1991 | full |

| Nr. | TABLE B - GIFT WRAP WITH MOON DESIGN (NL) | Year | Phase |
|---|---|---|---|
| 1 | Vroom en Dreesmann (folie-papier) | 2010 | last |
| 2 | Readshop "Finishing Touch"papier | 2010 | full |
| 3 | Kruidvat (Fa. Evora) | 2010 | last |
| 4 | Super De Boer | 2010 | first |
| 5 | Xenos | 2010 | first |
| 6 | Xenos | 2010 | last |
| 7 | Bart Smit | 2010 | first |
| 8 | Bart Smit | 2010 | last |
| 9 | Etos | 2010 | last |
| 10 | Bijenkorf (Fa. BIRI) | 2010 | last |
| 11 | Blokker | 2010 | full |
| 12 | Blokker | 2010 | last |
| 13 | Keijzer 002 (= Paco 002) | 2010 | last |
| 14 | Keijzer 005 | 2010 | last |
| 15 | Keijzer 032 | 2010 | last |
| 16 | Keijzer 017 en 022 | 2010 | first |
| 17 | Keijzer 90038 (= EPS/Sabe90038) | 2010 | last |
| 18 | Keijzer 90011 en 90013 (=Decoma/Sabe90011) | 2010 | first |
| 19 | BRUVO | 2010 | last |
| 20 | Bruna | 2010 | last |

| Nr. | TABLE C - SANTA CLAUS BOOKS (USA) | Authors | Publisher | Year | Phase |
|---|---|---|---|---|---|
| 1 | Night before Christmas | | Ideal's Christmas Classics | 2005 | full |
| 2 | Jingle Bells | Veronica Vasylenko (ill.) | Tiger Tales | 2007 | full |
| 3 | Christmastime is here, Little People Series | (Readers Digest) | Fisher Price | 2008 | last |
| 4 | Best Christmas book ever | R.Scarry | Sterling | 1978 | full |
| 5 | Night before Christmas (Scratch&Sniff) | Swanson (ill.) | Sterling | 2007 | full |
| 6 | Night before Christmas | Fujikawa (ill.) | Sterling | 2007 | full |
| 7 | Night before Christmas | Tom Browning (ill.) | Sterling | 2007 | full |
| 8 | Night before Christmas* | Eric Puybaret (ill.) | Imagine Books (NY) | | full |
| 9 | The Rooftop Hop (book + music) | Sheahan & Wright | Mr. Holidays Books | 2009 | first |
| 10 | Night before Christmas (record & story) | Tom Newsom (ill.) | Publications International | 2010 | full |
| 11 | Night before Christmas | Douglas Corsline (ill.) | Randomhouse | 1975 | full |
| 12 | Can you make Santa giggle? | | Ticcle'n Giggle Sound book | 2010 | first |
| 13 | Night before Christmas | Dana Regan (ill.) | Harper Festival | 1992 | full |
| 14 | Night before Christmas | Bruce Whatley (ill.) | Harper Collins Publishers | 1999 | full |
| 15 | Night before Christmas | Mary Engelbreit (ill.) | Harper Collins Publishers | 2002 | full |
| 16 | Flight of the Reindeer | Sullival & Wolff | Skyhore Publishing | 2010 | full |
| 17 | t Was the night before Christmas | Kat Whelan (ill.) | Tiger Tales | 2010 | first |
| 18 | The Toys Night before Christmas | Susanna Ronchi (ill.) | Templar Co. | 2008 | full |
| 19 | The Night before Christmas | Maggie Downer (ill.) | Tucker Slingsby | 2005 | full |
| 20 | Magic of Christmas (Treasury of Holiday Series) | (set of stories) | Little Tiger Press | 2005 | full |
| | idem, story #2 (in evening!) | idem | idem | idem | last |
| | idem, story #3 (in evening!) | idem | idem | idem | last |
| 21 | The Night before Christmas (1949) | Leonard Weisgard | Sandy Creek | 1997 | full |
| 22 | The night before Christmas (Little Golden Book) | Mircea Catusana (ill.) | Random House | 2001 | full |
| 23 | Here comes Christmas | Caroline Jayne Church | Scholastic | 2000 | first |
| 24 | Holly & Hall Moose "Our Christmas Adventure" | Clark et al. | Build-A-Bear Workshop | 2008 | full |
| 25 | The little fir tree | Brown & LaMarche | Harper Collins | 2005 | full |
| 26 | Night before Christmas | Tasha Tudor (ill.) | Little, Brown & Co. | 2009 | full |
| 27 | Night before Christmas - sticker book | Cathy Beylon | Dover | 1996 | full |
| 29 | The Night before Christmas in California | Smith & Egan | Gibbs-Smith | 1992 | full |
| | The Night before Christmas in California | Smith & Egan | Gibbs-Smith | 1992 | last |
| 29 | Christmas Fun - Where are they? | Tony Tallarico | Kidsbooks | 2010 | last |
| 30 | Merry Christmas - First look & find | Lobo & Sanboru | Publs. International (pi-books) | 2010 | first |

*also available with CD (Peter, Paul & Mary)

| Nr. | TABLE D - OTHER BOOKS | Authors | Publisher | Year | Phase |
|---|---|---|---|---|---|
| 1 | Clifford's Halloween | Norman Bridewell | Scholastic | 1986 | last |
| 2 | Trick or treat, little critter (Golden Book) | Gina & Mercer Mayer | Western Publ. Co. | 1993 | full |
| 3 | Halloween | Gail Gibbons | Holiday House (NY) | 1984 | last |
| 4 | Halloween Party | Feczko & Sims | Troll Assoc. (New Jersey) | 1985 | first |
| 5 | Great Pumpkin Charlie Brown | Schulz | Running Press (Philadelphia) | 2010 | full |
| 6 | Boo! Guess Who? (Peak-a-Board book) | Benrei Huang (ill.) | Random House | 1990 | full |
| 7 | Spooky, silly Halloween puzzles | Patrick Merrell | Sterling | 2010 | last |
| 8 | Trick or treat DORA! | (Children's Toy Store) | Simon & Schuster | 2010 | last |
| 9 | Angelina's Halloween | Katherine Holabird | American Girl | 2002 | full |
| 10 | The Witch who lives down the hall | Donna Guthrie | Hartcourt | 1985 | last |
| 11 | Who said BOO? (Halloween Poems) | Nancy White Carlstrom | Simon & Schuster | 1995 | full |
| 1 | Sint Maarten en het beertje | Schneider & Dusiková | De Vier Windstreken** | 2000 | last |
| 2 | Ik loop hier met mijn lantaren | Wilfel & Winterhager | Christofoor*** | 2007 | last |
| 3 | Sint Maarten liedjes (cover) | Wim Faber | Profiel, Bedum | 1989 | last |

*\*\*NordSüd Verlag, Zürich, 2000*
*\*\*\*Gabriel Verlag, Stuttgart, 2003*